\documentclass[11pt,a4paper]{article}

\usepackage{amsfonts}
\usepackage{amsmath}
\usepackage{amssymb}
\usepackage{mathtools}
\usepackage{graphicx}
\usepackage{cite}
\usepackage{color}

\usepackage[font=small,labelfont=bf,labelsep=period,justification=raggedright]{caption}

\makeatletter
\renewcommand*{\@biblabel}[1]{\hfill#1.}
\makeatother

\begin{document}

\begin{center}
{\Large\textbf {The mechanism of phagocytosis: two stages of engulfment}}
\\
\vspace{2mm}
David M. Richards$^{1,2,*}$,
Robert G. Endres$^{1,2}$
\\
\vspace{2mm}
$^1$ Department of Life Sciences, Imperial College, London, UK\\
$^2$ Centre for Integrative Systems Biology and Bioinformatics, Imperial College, London, UK\\
$^*$ E-mail: d.richards@imperial.ac.uk
\end{center}

\begin{abstract}
Despite being of vital importance to the immune system, the mechanism by which cells engulf relatively large solid particles during phagocytosis is still poorly understood. From movies of neutrophil phagocytosis of polystyrene beads, we measure the fractional engulfment as a function of time and demonstrate that phagocytosis occurs in two distinct stages. During the first stage, engulfment is relatively slow and progressively slows down as phagocytosis proceeds. However, at approximately half-engulfment, the rate of engulfment increases dramatically, with complete engulfment attained soon afterwards. By studying simple mathematical models of phagocytosis, we suggest that the first stage is due to a passive mechanism, determined by receptor diffusion and capture, whereas the second stage is more actively controlled, perhaps with receptors being driven towards the site of engulfment. We then consider a more advanced model that includes signaling and captures both stages of engulfment. This model predicts that there is an optimum ligand density for quick engulfment. Further, we show how this model explains why non-spherical particles engulf quickest when presented tip-first. Our findings suggest that active regulation may be a later evolutionary innovation, allowing fast and robust engulfment even for large particles.
\end{abstract}

%%%%%%%%%%%%%%%%%%%%
%%% INTRODUCTION %%%
%%%%%%%%%%%%%%%%%%%%

\section*{Introduction}

\begin{figure}[t!]
  \centerline{\includegraphics{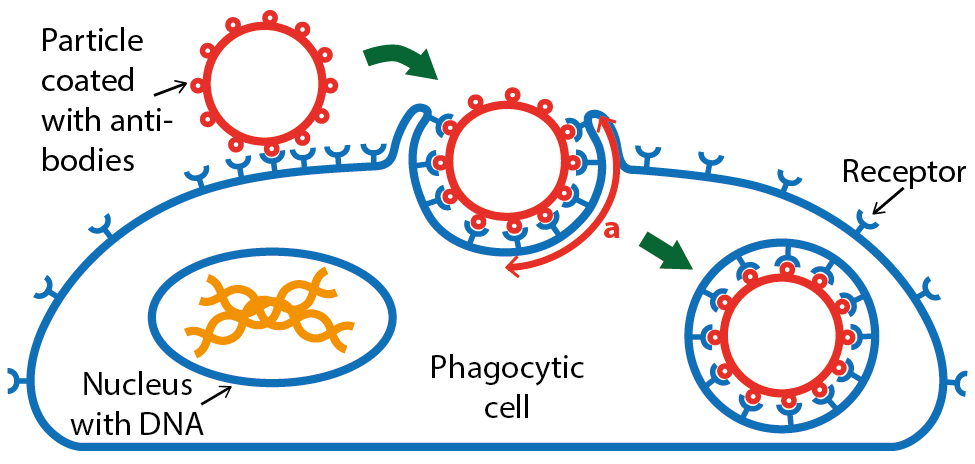}}
  \caption{{\bf Phagocytosis of a target particle.} Receptors on the cell surface bind ligand molecules on the target, such as a pathogen, dead cell or bead. As receptors bind more and more ligand molecules, the cell membrane progressively engulfs the target. Upon full engulfment, a phagosome is formed, which fuses with lysosomes, leading to digestion of the target. We denote the arc length of engulfed membrane by $a$, which gradually increases during engulfment.}\label{Fig1}
\end{figure}

Cells have evolved a whole host of mechanisms for ingesting particles and fluids. These vary from receptor-mediated endocytosis (absorption of small particles into clathrin-coated vesicles), to pinocytosis (the uptake of soluble material), to phagocytosis. Phagocytosis is the mechanism by which relatively large (over $0.5\rm{\mu m}$) particles, such as bacteria, dead cells or (as here) polystyrene beads, are internalized \cite{PhagocytosisReviewI,PhagocytosisReviewII,SwansonSignallingI}. During phagocytosis in immune cells such as neutrophils and macrophages, receptors in the cell membrane first recognise antibodies on the target, which causes membrane protrusions called pseudopodia to surround the target in a zipper-like mechanism (see Fig. \ref{Fig1}) \cite{GriffinI,GriffinII}. This is followed by fusion with lysosomes, acidification of the phagosome and degradation of the target.

The most widely-studied example of phagocytosis involves Fc$\gamma$ receptors, which recognise particles coated with immunoglobulin G (IgG) \cite{FcGammaReviewI,FcGammaReviewII}. Fc$\gamma$ receptors are expressed in white blood cells in four different classes, distinguished by their antibody affinity \cite{FcGammaReviewIII}. Upon binding to the Fc region of IgG, receptors signal via Syk kinases, small GTPases \cite{SrcAndSyk} and hundreds of other molecules, leading to substantial reorganization of the actin cytoskeleton and its contraction by multiple myosin isoforms \cite{SwansonReceptorDynamics,DartMyosin1G}. The resulting complexity \cite{UnderhillOzinsky} has the potential to obscure the fundamental processes and principles that, as often is the case in biology, may well be quite simple. In particular, the connection between fundamental physical mechanisms and biological regulation remains to be explained. Focusing on only the most important components, such as the receptors, ligand density and particle shape, may help elucidate the fundamental underlying mechanisms of engulfment.

Despite phagocytosis being discovered over a hundred years ago \cite{Metchnikoff}, there are relatively few theoretical models of cup formation and receptor dynamics. In \cite{Heinrich1B,HeinrichModel2}, finite element computations were used to argue that engulfment requires two crucial interactions: repulsion at the cup edge between the membrane and newly-polymerized actin, and a flattening force within the cup. Conversely, \cite{Howard} modeled the motion of receptors and F-actin in an attempt to explain the fact that phagocytosis normally either fully completes or stalls before half-engulfment. In \cite{TollisZipper} zipper-like engulfment was modeled as an actin-driven ratchet mechanism, leading to robust engulfment. However, none of these models allow for different physical and biological mechanisms to operate at different stages of engulfment.

In addition to these phagocytosis models, there is an elegant approach to modeling the related problem of endocytosis \cite{Freund}. This model maps the motion of receptors within the membrane to the well-known Stefan problem, in fact to the supercooled Stefan problem. The Stefan problem, introduced in the nineteenth century, applies to first-order phase transitions governed by the heat equation. The archetypal example is that of the melting of ice, where the boundary between water and ice continually moves as more ice melts. This is remarkably similar to the cup boundary during engulfment, which increases as receptors flow towards the cup, now determined by the diffusion equation. Using this correspondence, it was shown that there exists an optimum particle size during endocytosis corresponding to the shortest engulfment time: particles both larger and smaller than this optimum size take longer to engulf. Whilst phagocytosis is considered a more active, more regulated process than endocytosis, it is quite possible that early events in phagocytosis are driven by passive processes and may share similarities with endocytosis.

In this article we focus on the progression of engulfment, in particular on the rate of engulfment and its dependence on passive and active mechanisms. For a spherical bead with radius $R$ we define the engulfment, $a$, as the arc length from the centre of the cup to the edge of the cup. At the start of phagocytosis $a=0$, with $a$ then monotonically increasing during engulfment, reaching $\pi R$ upon full engulfment (see Fig. \ref{Fig1}). Naively, one might expect that $a$ initially increases quickly, as membrane wrinkles are used to extend the membrane around the particle, followed by a slower stage as new membrane must be synthesized or brought from internal stores \cite{TensionTwoPhases,NeutrophilWrinkles}. Interestingly, we will find exactly the opposite. By analysing multiple single-bead single-cell movies we find evidence for two distinct stages of engulfment: an initial slow stage followed by a much quicker second stage.

We then extend the passive endocytosis model of \cite{Freund} to describe phagocytosis of spherical beads. Since phagocytosis of large particles is considered more active (for example, more actin-dependent) than endocytosis, we then expand this model to include processes such as receptor drift and signaling, finding good agreement with the experimental data. We then examine the effect of ligand density on the engulfment time, predicting that particles with intermediate density are engulfed quickest. Finally, we study how ellipsoidal particles are engulfed, providing a potential explanation for why such particles are more likely to be engulfed when the highly-curved tip is presented to the cell first.

%%%%%%%%%%%%%%%%%%%%%%%%%%%%%%%%%%%%%%%%%%%%%%%%%%%%
%%% RESULTS: Phagocytosis proceeds in two stages %%%
%%%%%%%%%%%%%%%%%%%%%%%%%%%%%%%%%%%%%%%%%%%%%%%%%%%%

\section*{Results}

\subsection*{Phagocytosis proceeds in two stages}

\begin{figure}[t!]
  \centerline{\includegraphics[scale=1.2]{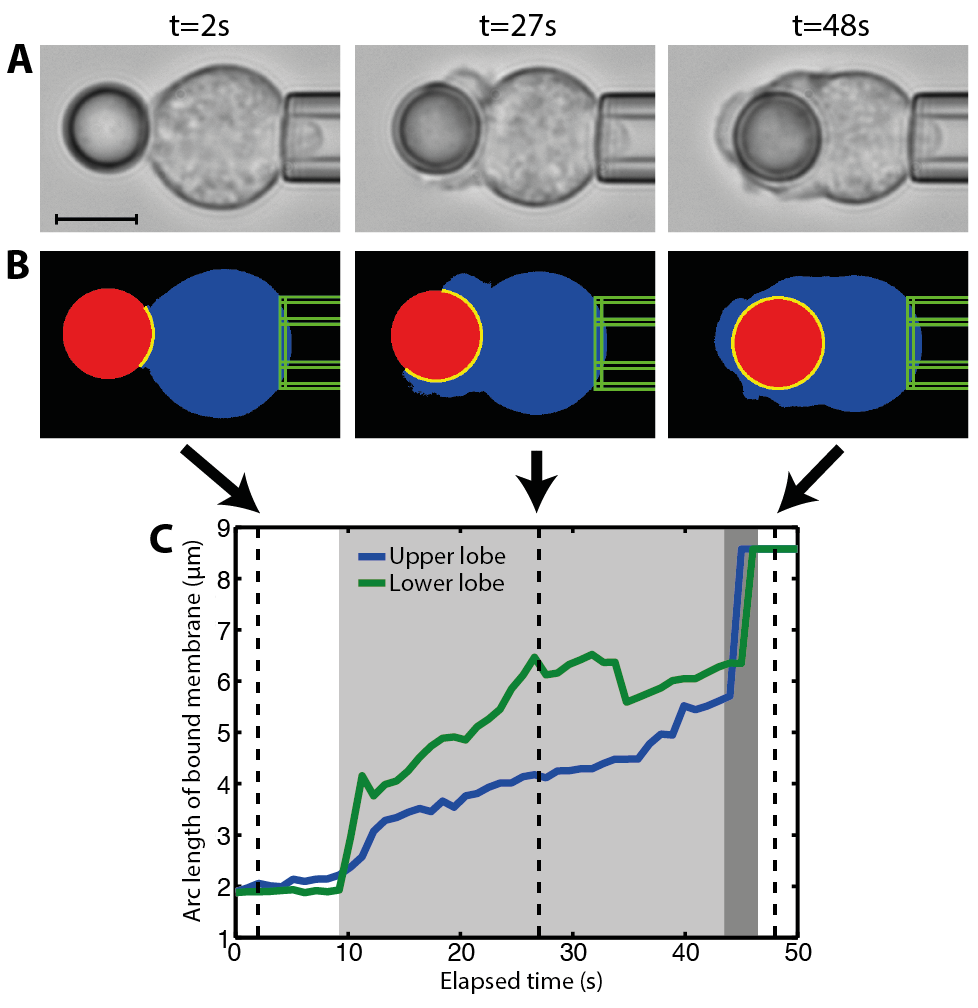}}
  \caption{{\bf Typical time-lapse movie and image analysis of neutrophil engulfing an IgG-coated bead.} Here the bead has diameter 4.6$\rm{\mu m}$. Data from \cite{Heinrich1B}. (A) Raw images of three frames at various stages of engulfment. At $t\approx2\rm{s}$ the bead has been released onto the cell, with a contact area of about $a\approx2\rm{\mu m}$. At this point engulfment has not yet started. At $t\approx27\rm{s}$ the bead is around half-engulfed, with the lower lobe noticeably ahead of the upper lobe. At $t\approx48\rm{s}$ engulfment is complete, the bead is entirely within the cell, and the phagosome is fully formed. Scale bar: 5$\mu\rm{m}$. (B) The same frames as in (A) after automatic image analysis, with the cell shown in blue, the bead in red and the outline of the pipette in green. The portion of the membrane attached to the bead is shown in yellow. (C) The engulfment as a function of time. For both the upper and lower lobes, after engulfment begins at around $t\approx10\rm{s}$, there is an initial slow stage (light gray) followed by a much quicker second stage (dark gray). Engulfment is complete by around $t\approx46\rm{s}$.}\label{Fig2}
\end{figure}

We analyzed six time-lapse movies of Fc$\gamma$R-mediated neutrophil phagocytosis, four with beads of $4.6\rm{\mu m}$ diameter and two with beads of $6.2\rm{\mu m}$ diameter (one of which is published in \cite{Heinrich1B}). Example frames are shown in Fig. \ref{Fig2}A. These were obtained by holding IgG-coated beads and neutrophils in separate micropipettes, before bead and cell were brought into contact and released. Image analysis was performed automatically in order to remove any human bias. Briefly, we used a combination of edge detection and Hough transforms to identify the position and size of the cell, bead and pipette. After removing the bead, a threshold was applied, allowing the shape of the membrane engulfing the bead to be identified, from which the arc length of membrane engulfing the cell was calculated. For full details see Methods. Examples of this analysis are shown in Fig. \ref{Fig2}B.

In our focal plane we can identify two lobes of the phagocytic cup, which we refer to as the top and bottom (Fig. \ref{Fig2}A). Although the top and bottom lobes are connected, we analysed them separately. Fig. \ref{Fig2}C shows a typical result of the engulfed arc length as a function of time. Initially there is a period (0-10s in Fig. \ref{Fig2}C) when the bead is in contact with the cell, although engulfment has not started. This results in a non-zero, approximately constant contact length between the bead and cell, due to adhesion between them. At some point ($\sim$10s in Fig. \ref{Fig2}C) engulfment begins and the engulfed arc length starts to increase as a function of time. Initially this engulfment is relatively slow, with the rate of further engulfment decreasing in time. However, at some point ($\sim$44s in Fig. \ref{Fig2}C) there is a sharp and rapid increase in engulfment rate, with full engulfment occurring soon after ($\sim$46s in Fig. \ref{Fig2}C). These two stages of engulfment, an initial slow stage followed by a rapid second stage, occurred for both $4.6\rm{\mu m}$ and $6.2\rm{\mu m}$ beads. Surprisingly, engulfment in the slow stage proceeds at different rates in the top and bottom lobes. This suggests that, at least initially, engulfment is a local process, depending only on the distribution of receptors at a given point on the membrane.

%%%%%%%%%%%%%%%%%%%%%%%%%%%%%%%%%%%%%%%%%%%%%%%
%%% RESULTS: Power laws describe engulfment %%%
%%%%%%%%%%%%%%%%%%%%%%%%%%%%%%%%%%%%%%%%%%%%%%%

\subsection*{Power laws describe engulfment}

\begin{figure}[t!]
  \centerline{\includegraphics[scale=1.2]{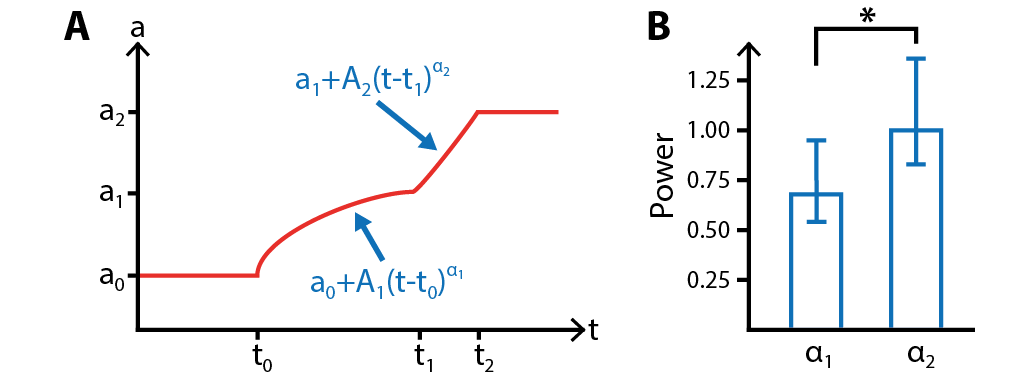}}
  \caption{{\bf Fitting to the general model.} (A) The general two-stage model is split into four regions. Initially, before engulfment begins, the contact length is constant with $a=a_0$. Second, after engulfment begins at $t=t_0$, the engulfed arc length is given as a power law with power $\alpha_1$ and prefactor $A_1$. Third, after $t=t_1$, the second step begins and $a$ is described by an independent power law with power $\alpha_2$ and prefactor $A_2$. Finally, after $t=t_2$ the particle is completely engulfed and $a=a_2$. (B) Comparison of the two powers, $\alpha_1$ and $\alpha_2$, showing the median and upper and lower quartiles. $n_{\alpha_1}=n_{\alpha_2}=12$ (6 beads $\times$ 2 lobes). A Mann-Whitney U test showed that the difference between $\alpha_1$ and $\alpha_2$ is significant ($p<0.02$, $U=107.5$).}\label{Fig3}
\end{figure}

To test the idea of two distinct stages, we fit our data to a general two-step model by assuming that each stage is described by an independent power law. To this end, we describe the engulfed arc length, $a(t)$, in four parts (Fig. \ref{Fig3}A).
\begin{enumerate}
  \item After the bead initially makes contact with the cell, there is a time when the bead sits on the surface of cell, with some non-zero engulfment, $a_0$.
  \item After some time $t_0$, engulfment begins and the first, slower stage starts. We describe this using a power law, $a(t)\sim A_1t^{\alpha_1}$, shifted so that $a=a_0$ at $t=t_0$.
  \item At time $t_1$ when the engulfment is $a_1$, the second, faster stage is initiated, described by an independent power law, $a(t)\sim A_2t^{\alpha_2}$, this time shifted so that $a=a_1$ at $t=t_1$.
  \item At time $t_2$, the bead is completely engulfed and the engulfed arc length reaches its maximum of $a_2$.
\end{enumerate}
Thus we model the engulfed arc length as
\begin{equation}
  \label{eq:gen_model}
  a(t) = \begin{dcases*}
           a_0 & for $t < t_0$ \\
           a_0 + A_1(t-t_0)^{\alpha_1} & for $t_0 \leq t < t_1$ \\
	           a_1 + A_2(t-t_1)^{\alpha_2} & for $t_1 \leq t < t_2$ \\
           a_2 & for $t \geq t_2$,
         \end{dcases*}
\end{equation}
where $a_1=a_0+A_1(t_1-t_0)^{\alpha_1}$ and $a_2=a_1+A_2(t_2-t_1)^{\alpha_2}$.

We observed that $a_0$, the initial contact between the bead and cell, is approximately constant in all our movies, even for different bead sizes, with an average and standard deviation of $a_0=2.10\pm0.20\mu\rm{m}$. Thus, to reduce the number of parameters in the general model, we set $a_0=2.10\mu\rm{m}$ from now on. Further, $a_2$, the arc length of membrane wrapping the bead at full engulfment, can readily be determined by examining the final few frames of each movie, after engulfment is complete. Thus, for any given movie, there are only six parameters to fit: the times $t_0$, $t_1$, the powers $\alpha_1$, $\alpha_2$, and the prefactors $A_1$, $A_2$ (see Methods for details).

The change from the slow to fast stage of engulfment occurs when the average fraction of bead engulfed is $0.47\pm0.10$, suggesting that the fast stage may be triggered around half-engulfment when further engulfment requires the purse-sting-like closure of the membrane around the bead \cite{SwansonContractile}. To examine the relative speeds of the first and second stages, we consider the fraction of total engulfment time spent in the first stage compared to that in the second, finding an average of $0.77\pm0.15$. This means that, on average, the first stage lasts over three times longer than the second stage, despite the fact that both stages engulf around half the total bead area. In addition to this, the parameters $\alpha_1$ and $\alpha_2$ give information about the exact manner in which engulfment proceeds during each stage. We find that the median of $\alpha_1$ is $0.68$ whereas that of $\alpha_2$ is $1.0$ (see Fig. \ref{Fig3}B for medians and upper and lower quartiles), suggesting that the stages may be governed by entirely different processes.

We can also examine whether there are differences in engulfment between our two bead sizes. For the average total engulfment time, $t_2-t_0$, we find, as expected, that the larger beads take longer to engulf ($37\pm8$s for $4.6\rm{\mu m}$ beads and $90\pm12$s for $6.2\rm{\mu m}$ beads). This relatively large increase is probably due to the available membrane becoming limiting for $6.2\rm{\mu m}$ beads, when the bead radius becomes similar to the cell radius. However, the fraction of bead engulfed at the start of the second stage is similar for both bead sizes ($0.49$ for $4.6\rm{\mu m}$ beads and $0.44$ for $6.2\rm{\mu m}$ beads). Our data is insufficient to determine whether this small difference is significant. Finally, the change in $\alpha$ between the stages is again consistent for both radii ($0.63\to1.0$ for $4.6\rm{\mu m}$ beads and $0.83\to1.1$ for $6.2\rm{\mu m}$ beads), suggesting that the two-stage mechanism is a general feature of phagocytosis, independent of bead size.

%%%%%%%%%%%%%%%%%%%%%%%%%%%%%%%%%%%%%
%%% RESULTS: Pure diffusion model %%%
%%%%%%%%%%%%%%%%%%%%%%%%%%%%%%%%%%%%%

\subsection*{Pure diffusion model}

\begin{figure}[t!]
  \centerline{\includegraphics[scale=1.2]{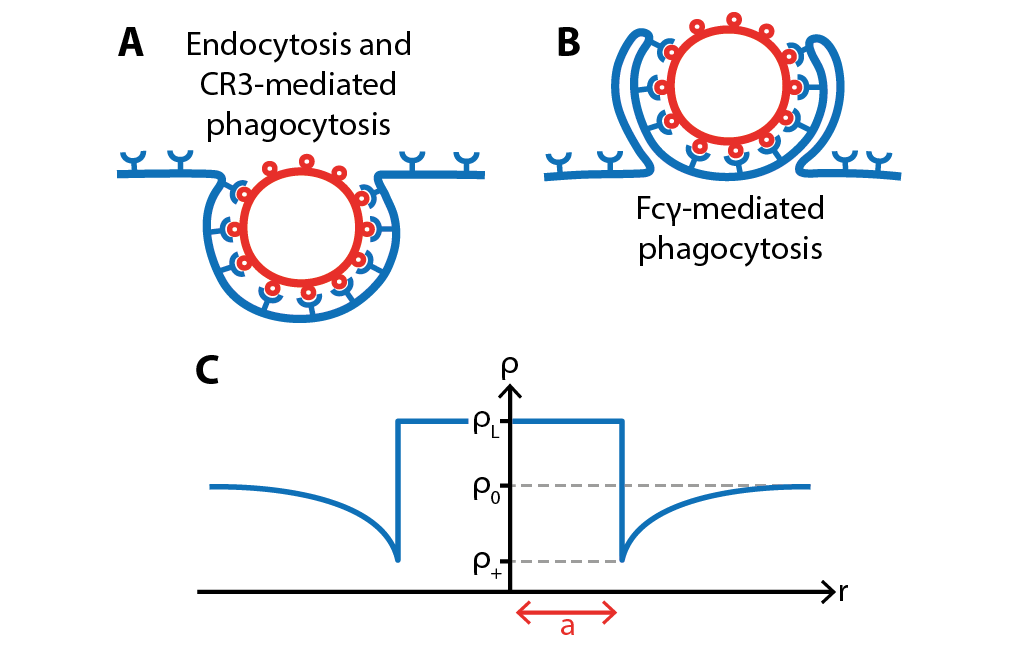}}
  \caption{{\bf Types of phagocytosis and the model variables.} (A) In CR3-mediated phagocytosis the particle sinks into the cell, in a manner similar to endocytosis. (B) In contrast, in Fc$\gamma$-mediated phagocytosis, the membrane extends via pseudopod extensions outwards around the particle. Despite these differences, we believe that the movement of receptors within the membrane is similar in both cases and can be described by the same simple model. (C) Sketch of the receptor density, $\rho$, which depends only on the distance from the centre of the cup, $r$, and the time, $t$. Within the cup, which has arc length $a$, the receptor density is always fixed at $\rho_L$, whereas the density at infinity, which is the same as the initial density, is $\rho_0$. At the edge of the cup, the receptor density is fixed at $\rho_+$, which is calculated by considering the free energy. In the pure diffusion model, $\rho_+$ is always less than $\rho_0$, so that receptors always flow in towards the cup and increase the cup size.}\label{Fig4}
\end{figure}

To try to understand the possible origins of such different dynamics, we consider simple models, focusing mainly on the motion of the Fc$\gamma$ receptors. Motivated by \cite{Freund}, we consider a circularly-symmetric two-dimensional membrane, with the origin the point at which the bead first touches the cell. This is a reasonable assumption given the spherical symmetry of the bead. The model describes aspects both of endocytosis \cite{Freund} and of CR3-mediated phagocytosis where the bead normally ``sinks'' into the cell as in Fig. \ref{Fig4}A. Even in Fc$\gamma$-mediated phagocytosis, where engulfment is by extension of the membrane around the bead (Fig. \ref{Fig4}B), the problem is still one of receptors moving in a two dimensional membrane. Since in our simple model we neglect the details of the cup shape, the only relevant difference is that the total curvature is greater in the Fc$\gamma$-mediated case (although the local curvature is approximately the same, a greater area of membrane is curved). Thus by including this extra curvature, we can also apply the same model to Fc$\gamma$-mediated phagocytosis.

Due to the assumed circular symmetry the problem reduces to receptors moving on a semi-infinite one-dimensional line. We parametrise this line by the distance from the origin, $r$, and describe the receptor density by $\rho(r,t)$, where $t$ is the time. Before contact with any bead, the receptor density is independent of $r$ and is given by $\rho_0$. During engulfment the receptor distribution is no longer constant, although the density at infinity is always $\rho_0$. The second variable is the cup size, $a(t)$, the engulfed arc length measured from the centre of the cup. Within the engulfed region, where $r<a$, receptors are attached to the ligands on the bead. As such their density is assumed to be related to the ligand density on the bead and, for simplicity, we assume that $\rho$ in this region is constant and given by $\rho_L$, where $\rho_L>\rho_0$. We denote the receptor density at the cup edge, $\rho(a(t),t)$, by $\rho_+(t)$. These variables and constants are shown in Fig. \ref{Fig4}C.

There are various possible physical mechanisms for the motion of the receptors in the non-engulfed region of the membrane, $r\ge a$. The first we consider is a purely passive process, where receptors simply diffuse around the membrane \cite{Freund}. Diffusion of $\rho$ is then described by $\frac{\partial\rho}{\partial t}=D\nabla^2\rho$, where $\nabla^2$ is the radial part of the Laplacian in cylindrical coordinates and $D$ is the diffusion constant. The evolution of $a$ is determined by conservation of receptors. Receptors move from outside to inside the cup region with flux $-D\rho'_+$, where $\rho'_+$ is shorthand for $\frac{\partial\rho}{\partial r}$ evaluated at $r=a$. This flux increases the density of receptors in the boundary region from $\rho_+$ to $\rho_L$, from which the rate of change of $a$ follows. Thus our pure diffusion model, for the non-engulfed region of the membrane where $r\ge a$, which is identical to the (supercooled) one-dimensional Stefan problem, is described by
\begin{subequations} \label{eq:diff_eqs}
  \begin{align}
    \frac{\partial\rho}{\partial t} &= \frac{D}{r} \frac{\partial}{\partial r}\left( r\frac{\partial\rho}{\partial r} \right), \\
    \frac{da}{dt} &= \frac{D\rho'_+}{\rho_L-\rho_+}, \label{eq:diff_eqs2}
  \end{align}
\end{subequations}
with initial conditions $\rho(r,0)=\rho_0$ and $a(0)=0$.

To find a solution to this system one extra condition is needed. Following \cite{Freund} we require that there is no free-energy change as receptors move into the engulfed region and the cup boundary is extended, indicating that all energy from receptor-ligand binding and configurational receptor entropy is used for bending the membrane and engulfment. We consider three contributions to the free energy: the binding between receptors and ligands, the curvature of the membrane, and the receptor entropy. The negative energy from binding is necessary to cancel the positive contribution from membrane curvature. We can write the free energy, $\mathcal{F}$, as
\begin{equation}
  \label{eq:free_energy}
  \frac{\mathcal{F}}{2\pi k_BT} = \int_0^a(-\rho_L\mathcal{E}+\tfrac{1}{2}\mathcal{B}\kappa_p^2)rdr + \int_0^\infty \rho\ln\left(\frac{\rho}{\rho_0}\right) rdr,
\end{equation}
where $\mathcal{E}$ is the binding energy per receptor-ligand bond, $\mathcal{B}$ is the bending modulus, and $\kappa_p$ is the radius of curvature of the bead. Although we do not explicitly include the membrane tension, such a term could be absorbed into the definition of $\mathcal{E}$. For a spherical bead of radius $R$, $\kappa_p=\frac{2}{R}$. Requiring no free-energy jump across the cup boundary implies that
\begin{equation}
  \label{eq:power_bal}
  \frac{\rho_+}{\rho_L} - \ln\left(\frac{\rho_+}{\rho_L}\right) = \mathcal{E} - \frac{\mathcal{B}\kappa_p^2}{2\rho_L} + 1,
\end{equation}
from which it follows that $\rho_+$, the receptor density at the cup edge, is a constant, independent of time. The (numerical) solution for $\rho_+$ gives the extra condition needed to uniquely solve Eqs. \ref{eq:diff_eqs}.

\begin{figure}[t!]
  \centerline{\includegraphics{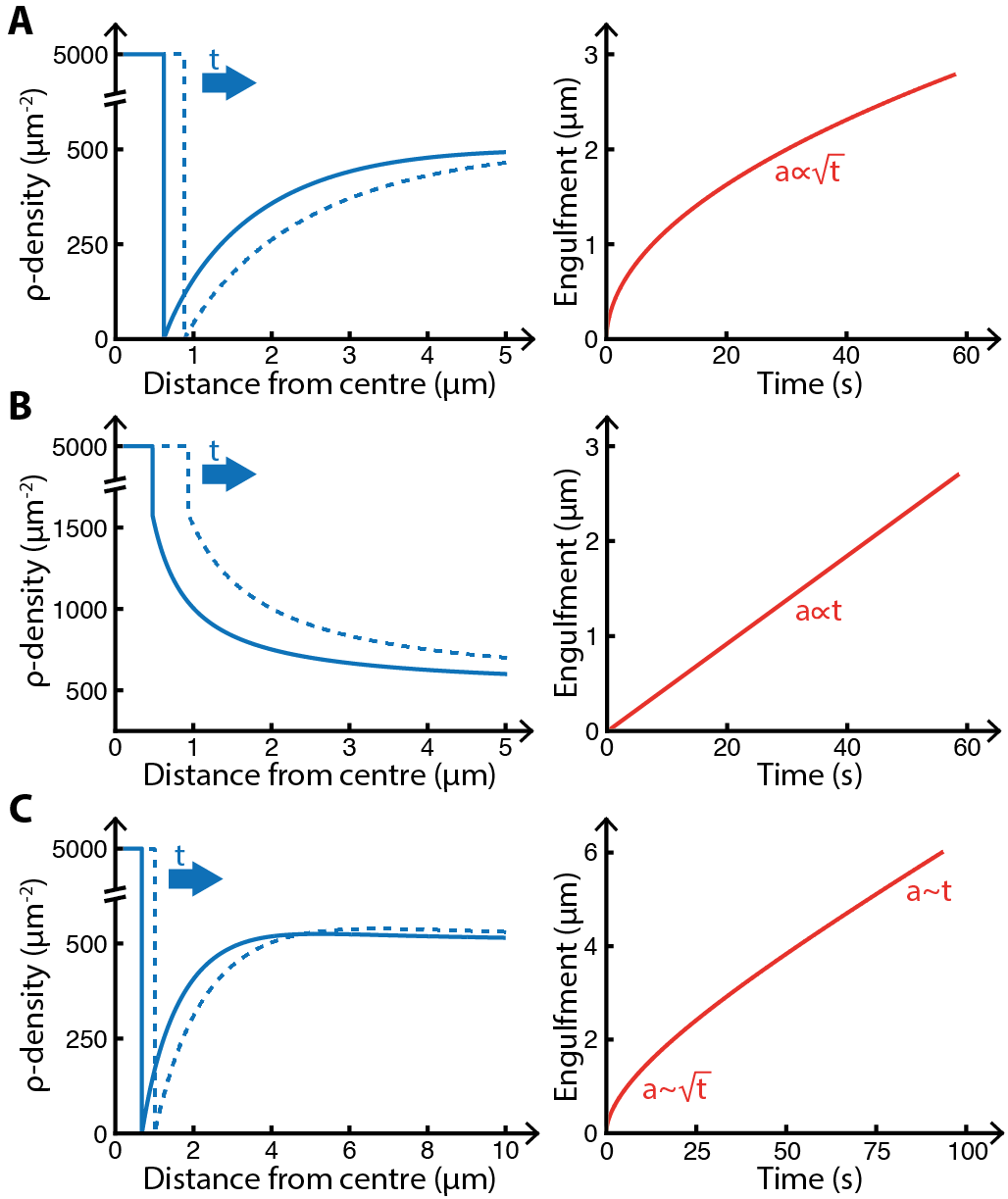}}
  \caption{{\bf Results from the pure diffusion, pure drift and diffusion-and-drift models.} Left: receptor density profile. Right: Engulfment against time. (A) Pure diffusion model with $D=1\mu\rm{m}^2\rm{s}^{-1}$. The receptor density, $\rho$, drops significantly just outside the cup so that $\rho_+<\rho_0$ and evolves to the right as engulfment proceeds. Note that the parameters are such that $\rho_+ \approx 0$. The engulfment increases as the square root of time. Receptor profile shown at $t=3\rm{s}$ (solid) and $t=6\rm{s}$ (dashed). (B) Pure drift model with $v=0.1\mu\rm{ms}^{-1}$. The receptor density has a completely different profile and decreases away from the cup, with $\rho_+$ now greater than $\rho_0$. Importantly, the engulfment now increases linearly in time. Receptor profile shown at $t=10\rm{s}$ (solid) and $t=20\rm{s}$ (dashed). (C) Diffusion and drift model with $D=1\mu\rm{m}^2\rm{s}^{-1}$ and $v=0.1\mu\rm{ms}^{-1}$. Engulfment now proceeds as a mixture of the pure diffusion and pure drift cases. Initially, when the receptor density is low, $a\sim\sqrt{t}$ as in the pure diffusion model. At later times, when the receptor gradient near the cup becomes approximately constant, $a$ becomes more linear in time and behaves more like the pure drift result. Receptor profile shown at $t=3\rm{s}$ (solid) and $t=6\rm{s}$ (dashed). Parameters: $\rho_0=50\mu\rm{m}^{-2}$, $\rho_L=5000\mu\rm{m}^{-2}$, $\mathcal{E}=15$, $\mathcal{B}=20$, $R=2\mu\rm{m}$, $L=50\mu\rm{m}$.}\label{Fig5}
\end{figure}

The solution, shown in Fig. \ref{Fig5}A, is then given by
\begin{subequations} \label{eq:diff_sol}
  \begin{align}
    \rho(r,t) &= \begin{dcases*}
              \rho_L & for $r<a$ \\
              \rho_0 - AE_1\left( \frac{r^2}{4Dt} \right) & for $r\ge a$
            \end{dcases*} \\
    a(t) &= 2\alpha\sqrt{Dt},
  \end{align}
\end{subequations}
where $E_1(x)=\int_x^\infty \frac{e^{-u}}{u}du$ is the exponential integral. The constant $\alpha$ is found by solving (numerically)
\begin{equation}
  \label{eq:diff_sol_alpha}
  \alpha^2 e^{\alpha^2} E_1(\alpha^2) = \frac{\rho_0-\rho_+}{\rho_L-\rho_+},
\end{equation}
after which $A$ is given by $A=(\rho_0-\rho_+)/E_1(\alpha^2)$. The most important property of the solution is the behavior of $a$, which increases as the square-root of time, $a\propto\sqrt{t}$. This is similar to the behavior we found when fitting the first stage of our movie data to the general model, suggesting that the first stage is controlled by passive receptor diffusion and capture.

%%%%%%%%%%%%%%%%%%%%%%%%%%%%%%%%%
%%% RESULTS: Pure drift model %%%
%%%%%%%%%%%%%%%%%%%%%%%%%%%%%%%%%

\subsection*{Pure drift model}
	
In contrast to this purely diffusive, passive model, we now consider the other extreme: a model where receptors are actively moved towards the edge of the cup. To do this we remove any diffusion and instead impose that the receptors drift with constant velocity towards the boundary $r=a$, requiring some active centripetal force, such as movement of receptors via retrograde actin flow \cite{YuRetrogradeFlow}. The movement of the receptors is now described by $\frac{\partial\rho}{\partial t}=v\nabla\cdot\rho$, where $\nabla\cdot$ only includes the radial part of the divergence in cylindrical coordinates and $v$ is the drift velocity. The equation for $a$ is also changed since now the flux across the cup boundary is $-v\rho_+$. Thus our second model, which is also a type of Stefan problem, but with diffusion replaced by drift, is given by
\begin{subequations} \label{eq:drift_eqs}
  \begin{align}
    \frac{\partial\rho}{\partial t} &= \frac{v}{r}\frac{\partial}{\partial r}\left( r\rho \right), \\
    \frac{da}{dt} &= \frac{v\rho_+}{\rho_L-\rho_+},
  \end{align}
\end{subequations}
with the same initial conditions as before.

Unlike for the diffusion model, there is no need for the extra condition obtained from free-energy considerations. The analytic solution, shown in Fig. \ref{Fig5}B, is given by
\begin{subequations} \label{eq:drift_sol}
  \begin{align}
    \rho(r,t) &= \begin{dcases*}
              \rho_L & for $r<a$ \\
              \rho_0\left( 1+\frac{vt}{r} \right) & for $r\ge a$
            \end{dcases*} \\
    a(t) &= \frac{1}{\sqrt{\rho_L/\rho_0}-1}vt.
  \end{align}
\end{subequations}
The receptor density at the edge is given by $\rho_+=\sqrt{\rho_L\rho_0}$, which means that, since Eq. \ref{eq:power_bal} will not in general be satisfied, it is impossible to avoid a free-energy jump across the boundary in this model. In contrast to the diffusion model, the rate of engulfment is now linear in time, $a\propto t$, which suggests that the drift model is more appropriate for the second, faster stage of engulfment, where, on average, we found $\alpha_2=1.0$.

%%%%%%%%%%%%%%%%%%%%%%%%%%%%%%%%%%%%%%%%%%%%%%
%%% RESULTS: Combining diffusion and drift %%%
%%%%%%%%%%%%%%%%%%%%%%%%%%%%%%%%%%%%%%%%%%%%%%

\subsection*{Combining diffusion and drift}

The pure diffusion and pure drift models are extreme cases, where the receptors either diffuse or drift, but not both. It is far more realistic to allow both behaviors, which results in a system that looks like a combination of Eqs. \ref{eq:diff_eqs} and \ref{eq:drift_eqs}. As for the pure diffusion model, we again consider the free energy and require that there is no energy jump across the cup. Despite the addition of drift, the resulting equation for $\rho_+$ is unchanged from Eq. \ref{eq:power_bal}, so that the receptor density at the cup is again constant, independent of time.

Unlike the pure diffusion and pure drift cases, this model cannot be solve analytically. Instead, as explained in Methods, we numerically solved the system on a finite-grid membrane. An example of the output is shown in Fig. \ref{Fig5}C. Near the cup the receptor profile looks similar to the pure diffusion case, with positive gradient so that diffusion (in addition to drift) causes receptors to flow across the boundary and increase the cup size. Further from the cup the density increases beyond the value at infinity, $\rho_0$, before turning round and decreasing back towards $\rho_0$. As with the pure diffusion model, the engulfed arc length $a(t)$ initially grows as $\sqrt{t}$. This is because, at early times, the flow of receptors into the cup is dominated by the relatively large gradient of $\rho$. At later times, when $\frac{\partial\rho}{\partial r}$ decreases, drift is the dominant cause of receptor flow, and $a(t)$ grows linearly with time. Thus the combined diffusion and drift model initially looks like pure diffusion, with later behavior more similar to pure drift.

%%%%%%%%%%%%%%%%%%%%%%%%%%%%%%%%%%%%%%%%%%%%%%%%%%%%%%%%%%%%%%%%%%%%%
%%% RESULTS: A full model that captures both phases of engulfment %%%
%%%%%%%%%%%%%%%%%%%%%%%%%%%%%%%%%%%%%%%%%%%%%%%%%%%%%%%%%%%%%%%%%%%%%

\subsection*{A full model that captures both phases of engulfment}

Although the diffusion and drift model displays $a\sim\sqrt{t}$ and $a\sim t$ regimes, it is clear that it cannot explain the sharp jump in Fig. \ref{Fig2}C. To progress further, we consider, in addition to the receptors, the role of signaling via a signaling molecule described by density $S$. Although phagocytosis in a real cell depends on many types of signaling molecules in complicated cascades, it is not unreasonable to assume that just one species will capture the essential manner in which signaling influences receptor dynamics. For simplicity we imagine that our signaling molecule only moves within the membrane, as happens for some small GTPases. Due to the circular symmetry of the system, the signaling molecule, as with the receptors, can be described as a function of the distance from the origin $r$ and the time $t$, so that $S=S(r,t)$.

\begin{figure}[t!]
  \centerline{\includegraphics[scale=1.2]{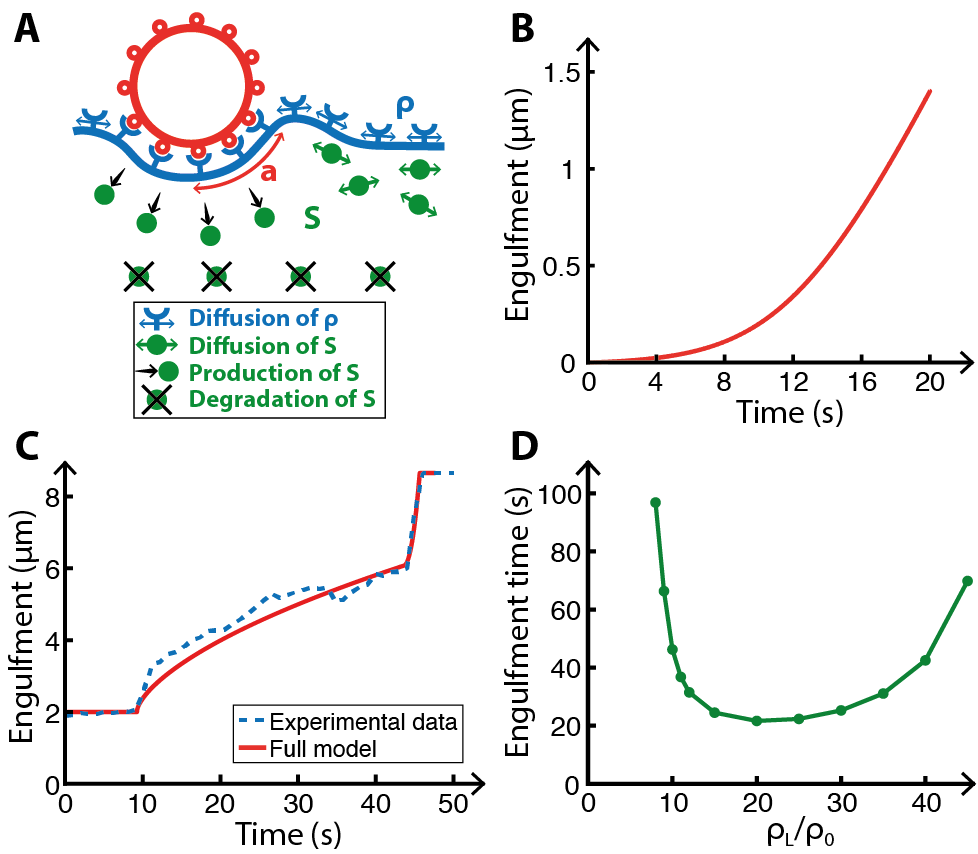}}
  \caption{{\bf Engulfment model with signaling.} (A) Sketch of the model which, in addition to the receptor density $\rho$, contains a signaling molecule with density $S$. Receptors can only diffuse (with diffusion constant $D$) and drift (with speed $v$), whereas the signaling molecule is produced within the cup with rate $\beta\rho_L$, degraded everywhere with lifetime $\tau$, and diffuses with diffusion constant $D_S$. (B) In contrast to other models, the rate of engulfment can now accelerate if the drift velocity depends linearly on $S$ via $v=v_1S$, tending to a constant as $t\to\infty$. Parameters: $v_1=20\mu\rm{m}^3\rm{s}^{-1}$, $\beta=0.1\rm{s}^{-1}$, $\tau=10\rm{s}$, $R=2\mu\rm{m}$, no diffusion. (C) In the full model, with a drift velocity that depends on the signaling molecule via a threshold, $S_0$, and an initial latent period, $t_0$, a sharp increase in engulfment rate can be obtained, which matches well with the measured data. Here the measured data is the average of the upper and lower lobes for the $4.6\rm{\mu m}$ bead shown in Fig. \ref{Fig2}C. Parameters: $D=3.8\mu\rm{m}^2\rm{s}^{-1}$, $v_1=6\mu\rm{m}^3\rm{s}^{-1}$, $S_0=0.498\mu\rm{m}^{-2}$, $\beta=0.4\rm{s}^{-1}$, $\tau=0.5\rm{s}$, $t_0=10\rm{s}$, $R=2.75\mu\rm{m}$. (D) The dependence of the full-engulfment time on the ligand density, $\rho_L$, showing a minimum at intermediate $\rho_L$. Parameters as in (C). Other parameters: $\rho_0=50\mu\rm{m}^{-2}$, $\mathcal{E}=3$, $\mathcal{B}=20$, $D_S=1\mu\rm{m}^2\rm{s}^{-1}$, $L=50\mu\rm{m}$.}\label{Fig6}
\end{figure}

We assume that initially there are no signaling molecules and that $S$ is produced with constant rate in response to receptor-ligand binding. As such, $S$ is only produced within the cup region where $r<a$. To counteract the production, we also allow $S$ to degrade everywhere with constant lifetime. Finally, we let $S$ diffuse throughout the membrane with diffusion constant $D_S$, which is generically different to the receptor diffusion constant $D$. Thus the dynamics of $S$ (as sketched in Fig. \ref{Fig6}A) are described by
\begin{align}
  \label{eq:sig_eq}
  \frac{\partial S}{\partial t} = \frac{D_S}{r} \frac{\partial}{\partial r}\left( r\frac{\partial S}{\partial r} \right) + \beta\rho_L\Theta(a-r) - \tau^{-1}S,
\end{align}
where $\Theta(x)$ is the Heaviside function, $\beta\rho_L$ is the binding rate, and $\tau$ is the lifetime.

In order that the signaling molecule can influence engulfment, {\it i.e.} that $S$ can actually signal, we need to couple $S$ to the dynamics of $\rho$. There are several ways we could do this, although we focus on the case where the density of the signaling molecule influences the drift speed. This can lead to radically different types of engulfment. For example, if we assume $v=v_1S$ for some constant $v_1$, it is possible to get accelerated cup growth as shown in Fig. \ref{Fig6}B. In such a model there is initially no signaling and so no drift. However, as the signaling molecule density at the cup edge increases, the drift velocity also increases, leading to quicker and quicker engulfment.

The effect of signaling on the drift velocity is likely to be co-operative, with multiple signaling molecules required to activate drift. Although we could model this via Hill-like behavior with a large Hill coefficient, we instead choose, for simplicity, to use a threshold $S_0$. Receptors diffuse until the signaling molecule at the cup edge reaches $S_0$. At this point, constant drift is turned on at all positions. This model, as shown in Fig. \ref{Fig6}C, contains a rapid increase in the rate of engulfment when $S_+$ reaches $S_0$.

To complete our model we introduce an initial latent period of length $t_0$, representing the period when the particle sits on the cell, before engulfment begins. By fitting the model parameters, we obtain good agreement with our experimental data. For example, Fig. \ref{Fig6}C shows the fit with the data from the $4.6\rm{\mu m}$ bead shown in Fig. \ref{Fig2}C (averaged over the upper and lower lobes). Although simple, our model correctly captures the main features we observed during engulfment, including the two-stage behavior of the cup edge.

%%%%%%%%%%%%%%%%%%%%%%%%%%%%%%%%%%%%%%%%%%%%%%%%%%%%%%%%%%%%%
%%% RESULTS: How engulfment depends on the ligand density %%%
%%%%%%%%%%%%%%%%%%%%%%%%%%%%%%%%%%%%%%%%%%%%%%%%%%%%%%%%%%%%%

\subsection*{How engulfment depends on the ligand density}

Using our full model, we now study the dependence of the engulfment time on the ligand density, $\rho_L$. This is not trivial since $\rho_L$ appears in the cup growth rate (Eq. \ref{eq:diff_eqs2}), the receptor density at the cup edge (Eq. \ref{eq:power_bal}), and the production rate of the signaling molecule (Eq. \ref{eq:sig_eq}). From our simulations we measure the full-engulfment time for various $\rho_L$ (Fig. \ref{Fig6}D).

Interestingly, the behavior is not monotonic. For small ligand densities, the signaling molecule production rate is small and $S$ only slowly increases towards $S_0$, leading to a long engulfment time. As $\rho_L$ increases, this production rate increases, which tends to reduce the engulfment time. However, at the same time, the cup growth rate decreases (since more ligands must be bound), so that the region of $S$ production ($r<a$) is smaller. This has the tendency to increase the engulfment time. These two competing effects lead to the non-monotonic behavior, with an initial decrease before a final rise as $\rho_L$ is increased. Notably, this predicts an optimum intermediate ligand density corresponding to the quickest possible engulfment. This is in sharp contrast to previous models (such as \cite{Freund} and \cite{Decuzzi2007}), where the engulfment time monotonically increases with increasing ligand density. However, such models do not include signaling and neglect the two-stage nature of engulfment.

Although some previous experimental work has addressed the dependence of phagocytosis on ligand density \cite{SwansonSignallingIII,SulchekLigandDensity}, this usually involved measuring the percentage of engulfed particles, rather than the progression of the cup with time. This is more a measure of how often engulfment stalls, and is unlikely to be directly related to the engulfment rate. The engulfment time was measured in \cite{SwansonSignallingIII}, with similar cup closure times at different $\rho_L$. However, since only two ligand densities were studied, this does not conflict with our prediction, for which ligand densities either side of the optimum density can have similar engulfment times.

%%%%%%%%%%%%%%%%%%%%%%%%%%%%%%%%%%%%%%%%%%%%%%%%%%%%%%%%%%%%%%%%%%%%%%%%%%%%%%%
%%% RESULTS: Ellipsoidal particles engulf quickest when presented tip-first %%%
%%%%%%%%%%%%%%%%%%%%%%%%%%%%%%%%%%%%%%%%%%%%%%%%%%%%%%%%%%%%%%%%%%%%%%%%%%%%%%%

\subsection*{Ellipsoidal particles engulf quickest when presented tip-first}

\begin{figure}[t!]
  \centerline{\includegraphics[scale=1.2]{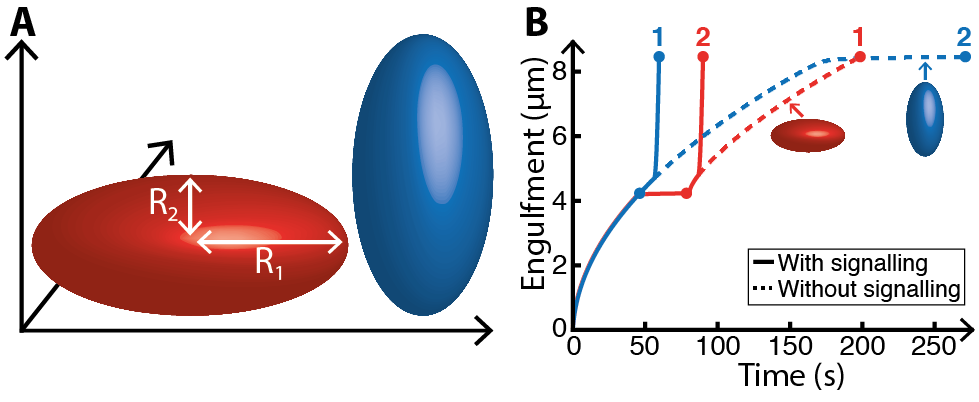}}
  \caption{{\bf Comparison of oblate and prolate spheroids.} (A) Sketch of the two spheroids, parameterised by $R_1$ and $R_2$. For oblate spheroids ($R_1>R_2$), the lowest curvature region is the first to be engulfed, whereas for prolate spheroids ($R_1<R_2$), the highest curvature region is engulfed first. (B) Progression of engulfment with time. Red: oblate spheroid with $R_1=R_2=4.1\mu\rm{m}$ and $R_3=0.62\mu\rm{m}$. Blue: prolate spheroid with $R_1=R_2=0.87\mu\rm{m}$ and $R_3=4\mu\rm{m}$. Solid lines: full model with signaling; dashed lines: pure diffusion model without signaling. Filled circles show half and full engulfment. Numbers show order of engulfment.}\label{Fig7}
\end{figure}

Finally, we study the effect of non-spherical particles. Since we are considering a one-dimensional model, our particles must be rotationally symmetric. For simplicity we only consider spheroids, that is ellipsoids where two of the principal axes have identical lengths. Thus spheroids are described by two parameters: $R_1$, the radius parallel to the membrane, and $R_2$, the radius perpendicular to the membrane (Fig. \ref{Fig7}A). The only difference required to the above model is that $\kappa_p$ in Eq. \ref{eq:power_bal} is now $2H$, where $H$ is the mean curvature of the spheroid. By using the standard parameterisation of the surface of a spheroid,
\begin{align}
  x =& R_1 \sin v \cos u, \nonumber \\
  y =& R_1 \sin v \sin u, \\
  z =& -R_2 \cos v, \nonumber
\end{align}
where $u\in[0,2\pi)$ and $v\in[0,\pi]$, we can write the mean curvature as \cite{EllipsoidCurvature}
\begin{equation} \label{eq:spheroid_curvature}
  2H = \frac{R_2\left[ R_1^2(1+\cos^2\!v) + R_2^2\sin^2\!v \right]}{\displaystyle R_1\left( R_1^2\cos^2\!v + R_2^2\sin^2\!v \right)^{3/2}}.
\end{equation}
At any given time, we determine the value of $v$ at the cup edge (corresponding to engulfed arc length $a$) and hence the curvature at $r=a$. This, via Eq. \ref{eq:power_bal}, gives the value of the receptor density at the cup edge, $\rho_+$. High curvature regions lead to relatively large values of $\rho_+$ and hence to slower engulfment.

Since we can only consider particles with two equal axes, we cannot directly compare the standing and lying-down versions of the same spheroid. Instead, we compare a spheroid where $R_1>R_2$ (which is an oblate spheroid as in the first shape in Fig. \ref{Fig7}A) with a spheroid with $R_1<R_2$ (which is a prolate spheroid as in the second shape in Fig. \ref{Fig7}A). To make the comparison meaningful, we ensure that both spheroids have the same circumference ({\it i.e.} the same total engulfment length). We also assume that these high-symmetry orientations are stable during engulfment.

To begin, we study the (unnatural) situation with signaling turned off, {\it i.e.} our pure diffusion model. For the oblate spheroid, the curvature is initially low, increasing to a maximum at half engulfment, before decreasing back to its initial value. This leads to a fast-slow-fast type of engulfment, with the slowest cup progression occurring near half engulfment (red dashed curve in Fig \ref{Fig7}B). In contrast, for the prolate spheroid, the highest curvatures occur at the beginning and end of engulfment. In principle this leads to slow-fast-slow engulfment, although the first stage is so short that it is better described as fast-slow engulfment (blue dashed curve in Fig \ref{Fig7}B). Since the highest curvature region occurs towards the end in the prolate case, when the engulfment rate is already low, this has a much greater effect on cup progression, meaning that (without signaling) the prolate spheroid takes the longer to engulf.

The situation, however, is markedly different for our full model with signaling (solid curves in Fig \ref{Fig7}B). Although the prolate spheroid takes longer than the oblate spheroid to reach full engulfment, it reaches half engulfment earlier. Consequently, the signaling molecule also reaches the threshold $S_0$ earlier, so that the second rapid engulfment stage begins sooner. Thus, with signaling, it is actually the prolate spheroid that is the first to complete engulfment. This agrees well with previous experimental observations, where non-spherical particles were found to be much easier to engulf when presented tip-first \cite{ParticleShape}. A similar effect was observed in \cite{Howard}, but in a model that ignores the two distinct stages of engulfment.

%%%%%%%%%%%%%%%%%%
%%% DISCUSSION %%%
%%%%%%%%%%%%%%%%%%

\section*{Discussion}

By analysing time-lapse movies of neutrophils exposed to IgG-coated beads, we found evidence of two distinct stages during phagocytosis engulfment. The first stage proceeds relatively slowly, taking, on average, three-quarters of the total engulfment time to reach only half-engulfment. In contrast, engulfment is much quicker in the second stage, often only taking a few seconds. This is perhaps opposite to the expected behavior: the initial availability of ``spare'' membrane within membrane wrinkles might be thought to lead to quick initial engulfment, with a subsequent slowing as new membrane must be created or brought from internal stores \cite{TensionTwoPhases,NeutrophilWrinkles}. However, this argument can be turned on its head: perhaps it is precisely the need to create new membrane at around half-engulfment that is the signal for the cell to enter a second, more active phase of engulfment, with a rapid increase in engulfment speed.

This experimental observation seems to conflict with the result in \cite{TensionTwoPhases}, which also found two stages, but with an initial rapid stage followed by a slower second stage. However, \cite{TensionTwoPhases} mostly involves macrophages spreading on a flat glass surface, whereas we study neutrophils engulfing spherical beads. This is likely to be substantially different both because of the huge difference in curvature (engulfment is highly dependent on curvature \cite{Freund,ParticleShape}) and, perhaps more importantly, because \cite{TensionTwoPhases} necessarily involves ``frustrated phagocytosis'', where engulfment can never complete. However, we believe the main difference occurs due to their large initial contact area, where cells placed on a glass surface instantly spread so that their initial contact area is around $140\mu\rm{m}^2$ (see their Fig. 1B), which is well above the total surface area for even our large $6.2\mu\rm{m}$ bead. Thus the cells in \cite{TensionTwoPhases} start with a contact area larger than that of complete engulfment for our beads, so that it seems likely that they may miss our initial slow stage when the cell is engulfing the first half of the target. Given this, the first stage in \cite{TensionTwoPhases} should be identified with our second stage. It is perhaps also relevant that \cite{TensionTwoPhases} used macrophages rather than neutrophils.

Motivated by the appearance of two stages, we then developed various models to describe engulfment. A complete mathematical model of phagocytosis is made difficult by the sheer number of parameters involved, which would in turn severely reduce its predictive power. Instead we made progress by considering simplified models that only focus on the key components, ignoring the role of myosins, the need for cytoskeletal remodelling, and the full three-dimensional membrane shape. Motivated by the elegant model of endocytosis in \cite{Freund}, our simplest model only focused on the Fc$\gamma$ receptors and assumed that the membrane appeared flat to the receptors. Although these are significant simplifications, we believe that our model still gives useful information about how the cell organises phagocytosis and can correctly capture the basic receptor dynamics. By the addition of receptor drift and a signaling molecule, we were able to develop a full, yet still simple, model that captures the sharp jump in engulfment rate.

The switch to the second stage occurs, on average, at almost exactly half-engulfment, which is precisely the point when the cell must start to tighten the cup around the top half of the particle for cup closure. This switch could provide an alternative explanation to the observation in \cite{Howard} that cells either fully engulf a target particle or stall before half way. If sometimes, perhaps due to insufficient signaling, the trigger for the second stage does not occur, this would leave cells at half-engulfment, without the necessary new membrane or modified receptor dynamics to enter the second stage and proceed to full engulfment.

How could the cell identify and signal the switch to the second stage of engulfment? Identification could be achieved in a variety of ways, perhaps by the increased membrane tension during engulfment \cite{TensionTwoPhases}, by the exhaustion of membrane wrinkles \cite{NeutrophilWrinkles}, or by the reduced speed of engulfment. With regards to signaling, there are various proteins that are known to be expressed at later stages of engulfment and may therefore serve as the signal for the switch \cite{SwansonSignallingI,SwansonSignallingIII,SwansonSignallingII}. For example, Rac1 and Rac2 (small signaling GTPases) localize to the cup well after engulfment begins, with Rac1 appearing first \cite{SwansonSignallingRac}. Similarly, PKC$\varepsilon$, a serine/threonine kinase, is only involved during the later stages of cup formation \cite{SignallingPKC}. Although originally thought to be important only when the phagosome is fully formed, it was shown in \cite{SwansonSignallingIII} that PKC$\varepsilon$ reaches half its maximum signal well before cup closure. Since stalled cups do not recruit PKC$\varepsilon$ \cite{SwansonSignallingIII} this would also tally with PKC$\varepsilon$ playing some role during the second stage of engulfment. Finally, phosphatidylinositol 3-kinase (PI3K) has been shown to be required for phagosome contraction during the latter stages of engulfment \cite{PI3KRole}.

The second stage is markedly faster than the first. With engulfment described by a power law, $a\sim At^\alpha$, this increased engulfment rate could be achieved either by increasing the prefactor $A$, increasing the power $\alpha$, or a combination of both. Our fit to the data (and our full model) suggests an increase in the power ($\alpha_2\gtrsim 1$), corresponding to the activation of drift in response to signaling (Fig. \ref{Fig6}). Such active drift motion of receptors towards the cup could be achieved, for example, by coupling the receptors to centripetal actin dynamics as observed in immunological synapses \cite{YuRetrogradeFlow}. In addition, it is likely that actin plays a role during the second stage, perhaps mediated by myosin IC \cite{SwansonContractile}. The full model leads to an interesting prediction for the receptor density near the cup. As seen in Figs. \ref{Fig5}A,B, the switch from diffusion to drift causes the sign of the receptor density gradient to switch immediately outside the cup: the depletion seen near the cup in the first stage disappears during the second stage. It would be interesting to tag the Fc$\gamma$ receptor and try to observe this. Further, and perhaps surprisingly, our model predicts that the quickest engulfment occurs for some intermediate value of the ligand density, a claim that would also be interesting to test. In addition, we found that our model can explain the observation that tip-first ellipsoids are easier to engulf. In fact, our model goes further and predicts the entire engulfment time course for such particles, a result that could also be tested, perhaps with the same micropipette assay used in \cite{Heinrich1B}.

Although our analysis favours an increase in the power $\alpha$ during the second stage, it is also possible that the cell implements the faster engulfment stage by increasing the prefactor $A$. This would still require active processes, but now the second stage would correspond to the pure diffusion model (as with the first stage), but with altered parameters. Such behavior could be achieved either by increasing the receptor diffusion constant $D$, increasing the receptor-ligand binding strength $\mathcal{E}$, decreasing the bending strength $\mathcal{B}$ or increasing the total number of receptors. Biologically these parameter changes could potentially be achieved by post-translational modifications of the receptors, by receptor clustering \cite{ReceptorClustering}, or by modifying the membrane structure (perhaps by altering lipid composition \cite{MembraneModificationI} or the connection with the actin cytoskeleton \cite{MembraneModificationII}). Models where the prefactor $A$ increases would differ from those where the power $\alpha$ increases since the receptor density would remain depleted just outside the cup even during the second stage. Thus, in principle, fluorescence microscopy of tagged receptors could distinguish between these two scenarios.

Understanding phagocytosis is of vital importance given its crucial role in the immune system and its relevance to drug delivery \cite{LinDrugDelivery}. Despite the staggering molecular complexity, simple physical mechanisms constrain and simplify the process, including membrane capacity \cite{MembraneCapacity}, particle stiffness \cite{ParticleStiffness} and particle shape \cite{ParticleShape}. Our simplified mathematical model shows that cells employ a multistage approach to engulfment, with radically different receptor dynamics occurring during different stages of engulfment. Potentially the second stage is a later evolutionary addition resulting in increased engulfment speed and robustness. Conversely, such an approach may help overcome the difficulty of engulfing the final part of the target, or perhaps function as a means of first examining the particle, a way of checking the desirability and feasibility of engulfment, before fully committing the cell machinery to the engulfment process.

%%%%%%%%%%%%%%%
%%% METHODS %%%
%%%%%%%%%%%%%%%

\section*{Materials and Methods}

\subsection*{Phagocytosis movies}
We studied six movies, four using $4.6\rm{\mu m}$-diameter beads and two using $6.2\rm{\mu m}$-diameter beads. Briefly, polystyrene beads were incubated first in bovine serum albumin (BSA) and then with rabbit anti-BSA antibody. Human neutrophils and beads were then picked up in separate micropipettes and brought into contact. Upon adhesion, the bead was released and phagocytosis observed under a bright field microscope using a 63$\times$ objective. The aspiration pressure was continually adjusted to ensure the length of cell within the micropipette was constant. Since only a small part of the cell is within the pipette, we believe that this continual pressure adjustment is unlikely to effect phagocytosis. This is confirmed by previous work that uses the same setup \cite{Heinrich1B,Zymosan,Heinrich0}. For details see \cite{Heinrich1A}.

\subsection*{Image analysis}
For each frame of a movie, the cell, bead and pipette were automatically identified. Initially edge detection, using the Sobel method, was performed, after which the bead was found used a Hough transform. This identified the centre and radius of the bead in each frame. The pipette was identified by searching for a set of horizontal lines, capped by a vertical line (as in Fig. \ref{Fig2}B). Due to the phagocytic cup, the cell is not well-described by a sphere. However, it was still possible to find the ``body'' of the cell (the cell minus the cup) by assuming the body was spherical. After removing the bead, pipette and cell body from each frame, all that remained is the phagocytic cup, which was separated from the background by imposing a threshold. Finally we measured the size of the phagocytic cup, defined as the arc length of bead circumference bound to the cell membrane.

\subsection*{Data fitting}
We scanned through a discretised version of parameter space that was described by $A_1,A_2 = \{0.05,0.1,0.15,\ldots,5.0\}$, $\alpha_1,\alpha_2 = \{0.05,0.1,0.15,\ldots,1.5\}$, $t_0,t_1 = \{0,1,2,\ldots,T\}$, where $T$ is the total length of the movie and where $t_0<t_1$. For each point in this space we define the error as $\sum_i\left(a_i-\bar{a}_i\right)^2$, where $a_i$ is the measured engulfment from frame $i$, $\bar{a}_i$ is given by Eq. \ref{eq:gen_model} applied at $t=t_i$, and the sum runs over all frames. The minimum error gives the best fit to the measured data. To test that the difference between the $\alpha_1$ and $\alpha_2$ distributions is statistically significant, we used the Mann-Whitney U test, a non-parametric test often used for non-normally distributed data, finding significance with $p<0.02$ ($U=107.5$, $n_{\alpha_1}=n_{\alpha_2}=12$).

\subsection*{Numerical simulations}
The diffusion-and-drift model and the model with signaling were solved numerically as follows. The membrane was represented by a finite grid of length $L=50\mu\rm{m}$ and spacing $\Delta r=0.01\mu\rm{m}$, labelled by $i$. Given the values of $\rho_i(t)$, $S_i(t)$ and $a(t)$ at some time $t$, the values at the next time step, $t+\Delta t$ ($\Delta t = 2.5\times10^{-5}\rm{s}$), were found from the Euler method and by imposing the boundary condition $\rho_{i(a)}(t+\Delta t)=\rho_+$, where $i(a)$ is $a$ expressed as an integer number of lattice steps and $\rho_+$ is found by solving Eq. \ref{eq:power_bal}. As a check of the numerical method and to ensure that values of $\Delta r$ and $\Delta t$ were sufficiently small, we checked that the numerical and analytic solutions matched for the pure diffusion and pure drift models. We also checked that decreasing $\Delta r$ and $\Delta t$ did not noticeably change the results, indicating convergence.

\subsection*{Parameter Values}
Parameter values were chosen as follows. $D=1\mu\rm{m}^2\rm{s}^{-1}$: close to Fc$\gamma$ receptor diffusion constant reported in \cite{Howard}. $v=0.1\mu\rm{ms}^{-1}$: chosen so that engulfment after one minute is similar in the pure diffusion and pure drift models. $\rho_0=50\mu\rm{m}^{-2}$: typical receptor density \cite{Freund,ReceptorDensity}. $\rho_L=500\mu\rm{m}^{-2}$: typical ligand density \cite{LigandDensity}. $\mathcal{E}=15$: measured Fc$\gamma$R-IgG binding free energy ($15k_BT$) \cite{BindingValue}. $\mathcal{B}=20$: typical value of bending modulus ($20k_BT$) \cite{BendingModulusValue}. $R=2\mu\rm{m}$: similar radius to the beads in our data. $L=50\mu\rm{m}$: approximate circumference of our cells.

%%%%%%%%%%%%%%%%%%%%%%%
%%% ACKNOWLEDGMENTS %%%
%%%%%%%%%%%%%%%%%%%%%%%

\section*{Acknowledgments}
We thank Volkmar Heinrich for providing movies of phagocytosis, and Benoit Raymond, Marianne Guenot and Gadi Frankel for useful discussions. DMR and RGE were supported by BBSRC grant BB/I019987/1. RGE also acknowledges funding from ERC Starting Grant 280492-PPHPI.

%%%%%%%%%%%%%%%%%%%%
%%% BIBLIOGRAPHY %%%
%%%%%%%%%%%%%%%%%%%%

\end{document}